\documentclass[12pt]{article}
\usepackage{amsfonts}
\usepackage{amssymb}
\usepackage[dvips]{graphicx}
\newcommand{\Eqn}[1]{&\hspace{-0.5em}#1\hspace{-0.5em}&}

\newcommand{\varth}{\vartheta}

\newcommand{\bbC}{{\mathbb C}}
\newcommand{\bbR}{{\mathbb R}}

\def\ep{\epsilon}

\def\pint#1 {- \!\!\!\!\!\!\!\! \,\int_{#1}}
\def\ni       {\noindent}

\def\comma      { \, , }
\def\period     { \, . }
\def\semiket#1  { \, #1 \, \rangle \, }
\def\del        {  \partial  }

\def\half       {  {1\over 2}  }
\def\abs#1      {  \, \vert #1 \vert \,   }
\def\Im#1    { \, {\rm Im } \, #1  }
\def\Re#1    { \, {\rm Re}  \, #1  }

\def\trans#1  { {}^{t} \! \vec{#1} }
\def\binom#1#2 { \vecii{ {}_{#1} }{\raisebox{.5ex}{$ {}^{#2} $}} }
\def\sqbinom#1#2 { \Bigl(\begin{array}{c} {}_{#1}
                       \\ \raisebox{.5ex}{${}^{#2}$} \end{array}\Bigr)^2  }

\def\r12    {\frac{r_1}{r_2}}

\def\calN   {{\cal N}}

\def\sn    { {\rm sn}  }
\def\cn    { {\rm cn} }
\def\dn    { {\rm dn} }
\def\ns    { {\rm ns} }
\def\sc    { {\rm sc} }

\def\nc    { {\rm nc} }
\def\ds    { {\rm ds} }
\def\sd    { {\rm sd} }
\def\cd    { {\rm cd} }
\def\dc    { {\rm dc} }

\def\phiinv   { \varphi^{\sigma} }
\def\rinv   { r^{\sigma} }
\def\vecii#1#2      {  { #1 \choose #2 }  }
\def\veciii#1#2#3   {  \left(\begin{array}{c}#1\\#2\\#3\end{array}\right)  }
\def\matrixii#1#2#3#4            {  \Bigl( \begin{array}{cc}#1&#2\\#3&#4
                                       \end{array} \Bigr) }
\def\matrixiii#1#2#3#4#5#6#7#8#9 {  \left(\begin{array}{ccc}#1&#2&#3\\
                                     #4&#5&#6\\#7&#8&#9\end{array}\right)  }
\def\eqb         {  \begin{eqnarray}  }
\def\eqe           {  \end{eqnarray}  }
\def\nn               {  \nonumber  }
\def\sectionnumbering { \setcounter{equation}{0}
         \renewcommand{\theequation}{\arabic{section}.\arabic{equation}}}
\def\appendixnumbering { \setcounter{equation}{0}
         \renewcommand{\theequation}{\Alph{section}.\arabic{equation}}}
\def\msection#1{ \addtocounter{section}{1} \setcounter{subsection}{0}
                 \sectionnumbering
   \par \bigskip
      \par \bigskip \noindent
   {\bf \arabic{section} \quad  #1 }
    \par \bigskip}
\def\appsection#1{\addtocounter{section}{1} \setcounter{subsection}{0}
                 \appendixnumbering
      \par \bigskip \par \bigskip \noindent
   {\bf   #1 }
    \par \bigskip}
\def\msubsection#1{\addtocounter{subsection}{1}
      \par \bigskip \noindent  {\normalsize\it
      \arabic{section}.\arabic{subsection} \quad #1  }
   \par \medskip }

\def\csectionast#1    { \begin{center}
    {\large\bf #1  }   \end{center} \par \bigskip}
\def\titleandfile#1#2   {  \begin{center}{\large\bf #1}\end{center}
                            \par\begin{flushright} #2 \end{flushright} 
                          }
\renewcommand{\thefootnote}{\fnsymbol{footnote}}

\newenvironment{namelist}[1]{%
   \begin{list}{}
      {
       \settowidth{\labelwidth}{#1}
       \setlength{\leftmargin}{1.1\labelwidth}}
}{%
 \end{list}}

\begin{document}
\def\papertitlepage{\baselineskip 3.5ex \thispagestyle{empty}}
\def\preprinumber#1#2#3{\hfill \begin{minipage}{2.6cm} #1
                \par\noindent #2
              \par\noindent #3
             \end{minipage}}
\renewcommand{\thefootnote}{\fnsymbol{footnote}}
%
%
\papertitlepage
\setcounter{page}{0}
\preprinumber{}{UTHEP-588}{arXiv:0907.5259}
\baselineskip 0.8cm
\vspace*{2.0cm}
\begin{center}
{\large\bf\mathversion{bold} A note on string solutions in $AdS_{3}$}
\end{center}
\vskip 4ex
\baselineskip 1.0cm
\begin{center}
        { Kazuhiro ~Sakai\footnote[2]{\tt sakai@phys-h.keio.ac.jp}, } \\
 \vskip -1ex
    {\it Department of Physics, Keio University}
 \vskip -2ex   
    {\it Hiyoshi, Yokohama 223-8521, Japan} \\
 \vskip 2ex
     { Yuji  ~Satoh\footnote[3]{\tt ysatoh@het.ph.tsukuba.ac.jp}}  \\
 \vskip -1ex
    {\it Institute of Physics, University of Tsukuba} \\
 \vskip -2ex
   {\it Tsukuba, Ibaraki 305-8571, Japan}
\end{center}
\vskip 10ex
%
\baselineskip=3.5ex
\begin{center} {\bf Abstract} \end{center}

\par\medskip
\ We systematically search for classical open string solutions in $AdS_{3}$ within 
the general class expressed by elliptic functions (i.e., the genus-one finite-gap solutions).
By explicitly solving the reality and Virasoro conditions, we give a classification of 
the allowed solutions. When the elliptic modulus degenerates, we find a class of solutions 
with six null boundaries, among which two pairs are collinear. By adding the $S^{1}$ 
sector, we also find four-cusp solutions with null boundaries
expressed by the elliptic  functions.

%
%
%
%
%
%

\vspace*{\fill}
\ni
July 2009

\newpage
\renewcommand{\thefootnote}{\arabic{footnote}}
\setcounter{footnote}{0}
\setcounter{section}{0}
\baselineskip = 3.3ex
\pagestyle{plain}
\sectionnumbering
\msection{Introduction}

Classical open string solutions in the anti-de Sitter (AdS) space with null boundaries 
give  the scattering amplitudes in planar $\calN =4$ super Yang-Mills theory
at strong coupling \cite{Alday:2007hr} (for a review, see for example, \cite{Alday:2008yw}).  
Because of this, the problem 
of finding such solutions have attracted much attention 
\cite{Ryang:2007bc}-\cite{Dorn:2009kq}.
Though the solution with four null boundaries and cusps  is 
found in \cite{Kruczenski:2002fb,Alday:2007hr}, finding the solutions with more than 
four cusps is still challenging. Recently, 
a prescription to construct multi-cusp 
solutions is provided in \cite{AM2}.  
There, it is also discussed how to compute
the scattering amplitudes without using explicit form of the solutions,
and this is demonstrated in the case of the eight-cusp solutions.
Regarding the numerical multi-cusp solutions, see \cite{Dobashi:2008ia}.

With applications to the  scattering amplitudes in mind, 
we discuss the classical open string solutions in $AdS_3$. 
For this purpose, a good starting point would be a 
general construction of the classical string solutions in $dS_{2n+1}$
\cite{Krichever}, where the solutions are 
expressed by theta functions
and integrals defined over the underlying spectral curve.
This construction can  also be applied 
to $AdS_{2n+1}$.\footnote{
For the closed strings in $AdS_{3} \times S^{1}$, another general construction  is given
in \cite{Kazakov:2004nh}. 
Explicit genus-one finite-gap 
solutions are discussed in \cite{Hayashi:2007bq}.
}
However,
for constructing relatively simple solutions,
it may be easier to make an ansatz of the finite-gap form
(i.e., the general form implied by \cite{Krichever}), 
where one regards the periods and the integrals as
free paramters, and search for particular solutions which 
satisfy definite reality, Virasoro and boundary conditions.

In this paper, 
we take this approach for the genus-one finite-gap solutions (elliptic
solutions).
We determine the parameters of the solutions by explicitly 
solving the equations of motion, and the reality and Virasoro conditions.
As a result, we give a classification of the allowed genus-one finite-gap solutions.
 When the elliptic modulus degenerates, 
we also find a class of  solutions with six null boundaries, 
 among which two pairs are collinear.
The solutions are expressed simply by hyperbolic
and  exponential functions, and describe non-flat minimal surfaces
in $AdS_{3}$. The analysis can be generalized 
to the classical string solutions in $AdS_{5} \times S^{5}$. By adding $S^{1}$, as a
simple example, we find four-cusp solutions with null boundaries 
expressed 
by elliptic functions. 

The rest of this paper is organized as follows. In section 2, starting with the genus-one
finite-gap form, we solve the equations of motion and the normalization  condition.
We then  summarize the Virasoro condition and the reality condition.
By solving these conditions, we determine the allowed solutions and give
a classification in section 3. In section 4, we discuss examples of the solutions.
In particular, we present a class of  solutions with six null boundaries. 
In section 5, we analyze the case of the strings in $AdS_{3} \times S^{1}$, 
and find  four-cusp solutions expressed by the elliptic
functions. We conclude with a discussion in section 6. 
The appendix includes
our conventions and some formulas of the elliptic theta functions.

\msection{Genus-one finite-gap solutions}

We begin with parametrizing  the $AdS_3$ target space by the embedding coordinates
in $R^{2,2}$, namely,
$Y_a(\sigma_+,\sigma_-)$, $a=-1,0,1,2$, 
with a constraint
\eqb
\vec{Y}\cdot\vec{Y} :=
  -Y_{-1}^2-Y_0^2+Y_1^2+Y_2^2=-1 \period\label{Ynorm}
\eqe
They satisfy the equations of motion
\eqb
\partial_+\partial_-\vec{Y}
-(\partial_+\vec{Y}\cdot\partial_-\vec{Y})\vec{Y}=0 \comma 
 \label{Yeom}
\eqe
and the Virasoro constraints
\eqb
(\partial_\pm\vec{Y})^2=0 \period \label{YVirasoro}
\eqe
The solutions span minimal surfaces in $AdS_3$.
In the following, we concentrate on the Euclidean world-sheet with
$(\sigma_{+})^{*} = \sigma_{-}$. The case of the Lorentzian
world-sheet can be  discussed similarly.

To find the solutions, we introduce the vector 
$\vec{\varphi} = (\varphi_{1}, \phiinv_{1}, \varphi_{2}, \phiinv_{2})$ which satisfies
\eqb
1 \Eqn{=} \sum_{j=1}^2\varphi_j\varphi^\sigma_j\comma \label{varphinorm} \\
 0 \Eqn{=}(\del_{+} \del_{-} + u)\vec{\varphi} \comma \label{varphieom}\\
  0 \Eqn{=} \sum_{j=1}^2\del_{\pm} \varphi_j\del_{\pm}\varphi^\sigma_j 
 \comma \label{varphiVirasoro}
\eqe
with the self-consistent potential
\eqb
  u= \frac{1}{2}\sum_{j=1}^2\left(
    \del_{+} \varphi_j\del_{-} \varphi^\sigma_j
   +\del_{-}\varphi_j\del_{+}\varphi^\sigma_j
\right). \label{potu}
\eqe
The equations (\ref{varphinorm})-(\ref{varphiVirasoro}) are
equivalent to  (\ref{Ynorm})-(\ref{YVirasoro})
under the identification $\varphi = Y$, where
\eqb
   \varphi := \Biggl(\begin{array}{cc} \varphi_{1} & \varphi_{2} \\ -\phiinv_{2} & 
   \phiinv_{1}\end{array}\Biggr) \comma \quad
   Y := \Biggl(\begin{array}{cc} Y_{-1}+Y_{2} & Y_{1} + Y_{0} \\ Y_{1} - Y_{0} & 
    Y_{-1} -Y_{2}\end{array}\Biggr) \period
\eqe
As discussed shortly, 
more general identifications between $\varphi$ and $Y$
are possible.

In the genus-one case, the finite-gap solution
to  (\ref{varphinorm})-(\ref{varphiVirasoro})  takes 
the form  \cite{Krichever}
\eqb
\varphi_j\Eqn{=}r_j\frac{\varth_3(X_0)\varth_0(X+A_j)}
  {\varth_3(X_0+A_j)\varth_0(X)}e^{p_j^+\sigma_+ +p_j^-\sigma_-}, \nn \\
\varphi^\sigma_j\Eqn{=} \rinv_{j} \frac{\varth_3(X_0)\varth_0(X-A_j)}
  {\varth_3(X_0-A_j)\varth_0(X)}e^{-(p_j^+\sigma_+ +p_j^-\sigma_-)} \period
\eqe
Here, 
\eqb
 X \Eqn{=} U^+\sigma_+ + U^-\sigma_- + X_0 - K(k) \comma \label{X0X}
\eqe
and $K(k)$ is the complete elliptic integral of the first kind
with $k$ the elliptic modulus. 
$\varth_{a}(z)$ are the elliptic theta functions which have the quasi-periods 
$(2K(k),2iK'(k))$ with 
$K'(k) = K(k')$ and $(k')^{2} = 1-k^{2}$. 
Compared with the standard notation, we have rescaled the argument 
of the theta functions by $2K$. With this convention, for example, 
$\varth_{0}(z+K) = \varth_{3}(z)$ and 
$\sn\,z=\varth_3(0)\varth_1(z)/\varth_2(0)\varth_0(z)$. 
To make the following expressions simpler, we have shifted $X$ by
$K$ as in (\ref{X0X}), which results in the combination of $\varth_3$
and $\varth_0$ in ${\varphi}$.
Our conventions of the elliptic theta functions are summarized in
the appendix.

Other parameters should be determined by imposing appropriate 
conditions. 
First, one finds that  the normalization condition (\ref{varphinorm}) gives
\eqb
r_1 \rinv_{1}=\frac{\sn^2 A_2(1-k^2\sn^2A_1\,{\rm cd}^2 X_0)}{\sn^2 A_2-\sn^2 A_1},
\quad
r_2 \rinv_{2}=\frac{\sn^2 A_1(1-k^2\sn^2A_2\,{\rm cd}^2 X_0)}{\sn^2 A_1-\sn^2 A_2}
\comma \label{rr}
\eqe
for $A_{1} \neq A_{2}$. 
The case of $A_1 = A_2 $ is discussed later in section 3.3.\footnote{
In addition, when $A_{j} = 0, iK'$, some of the expressions below become singular.
In the case of $A_{j} = 0$,  the solution becomes of the exponential type  without
the theta functions.
The case with $A_{j} = iK'$ is treated as a limiting case
from $A_{j} \neq iK'$.}
In deriving this, we have used 
\eqb
\varphi_j\varphi_j^\sigma
= r_j \rinv_j \frac{1+(kk')^2 \, \sd^2(X+K)\,\sd^2 A_j}{1+(kk')^2 \, \sd^2 X_0 \,\sd^2 A_j}
= r_j \rinv_j \frac{1-k^{2} \sn^{2} A_{j} \, \sn^{2}X}{1-k^{2} \sn^{2} A_{j} \, \cd^{2}X_{0}}
\comma \label{phiphi}
\eqe
which follow from product identities of $\varth_a$.

Next, to consider the equations of motion,
we introduce 
\eqb
\beta_j^\pm :=Z(A_j)+\frac{p_j^\pm}{U^\pm}, \label{betadef}
\eqe
where $Z(z):=\partial_z\ln\varth_0(z)$.
With the help of the formula (\ref{addZ}), 
one then obtains
\eqb
\frac{\partial_+\partial_-\varphi_j}{\varphi_j}
\Eqn{=}U^+U^-\left[
\left(k^2\sn\,A_j\,\sn\,X\,\sn(X+A_j)-\beta_j^+\right)
\left(k^2\sn\,A_j\,\sn\,X\,\sn(X+A_j)-\beta_j^-\right)
\right.\nn\\
&&\hspace{4em}\left. -k^2\sn^2(X+A_j)+k^2\sn^2 X\right] \comma 
\eqe
and similar equations for $\phiinv_{j}$ with $A_{j}, p^{\pm}_{j}$ replaced
by $-A_{j}, -p^{\pm}_{j}$. For these to be equated with $-u$, the $X$-dependence
should be common to all $\varphi_{j},\phiinv_{j}$. This requirement fixes 
$\beta_j^\pm$ as
\eqb
\beta_j^+ + \beta_j^- \Eqn{=} -\frac{2\cn\,A_j\,\dn\,A_j}{\sn\,A_j}, \nn\\
\beta_j^+ \beta_j^- \Eqn{=} k^2\sn^2A_j+u_0, \label{betaeom}
\eqe
where $u_{0}$ is a constant. Substituting these, one obtains
\eqb
\frac{\partial_+\partial_-\varphi_j}{\varphi_j}
=U^+U^-(2k^2\sn^2X+u_0). \label{-u}
\eqe
On the other hand, the potential $u$ in (\ref{potu})
is evaluated using  the equations for $\varphi_j \phiinv_j $  in  (\ref{phiphi}). 
Some computations show that $-u$ is indeed given by the right-hand side of
(\ref{-u}), which verifies the equations of motion.

Third, let us turn to the Virasoro condition. Again, after some algebra, 
one finds that the constraint
\eqb
\sum_{j=1}^2
\Bigl( \frac{U^{-}}{U^{+}}\partial_+\varphi_j\partial_+\varphi^\sigma_j
+ \frac{U^{+}}{U^{-}}\partial_-\varphi_j\partial_-\varphi^\sigma_j \Bigr)
=0
\eqe
determines the constant $u_0=-u(X = 0)/U^{+}U^{-}$ to be
\eqb
u_0=2\left(\frac{1}{\sn^2 A_1}+\frac{1}{\sn^2 A_2}-1-k^2\right) \comma
\eqe
whereas the other  constraint reads
\eqb
0\Eqn{=}
\sum_{j=1}^2
\Bigl( \frac{U^{-}}{U^{+}} \partial_+\varphi_j\partial_+\varphi^\sigma_j
- \frac{U^{+}}{U^{-}}\partial_-\varphi_j\partial_-\varphi^\sigma_j \Bigr)  \nn \\
\Eqn{=}
2U^+U^-
\frac{\sn^2 A_1\,\sn^2 A_2}{\sn^2 A_2-\sn^2 A_1}
\sum_{j=1}^2(-)^{j+1}\frac{\cn\,A_j\,\dn\,A_j}{\sn^3 A_j} (\beta^{+}_{j} - \beta^{-}_{j})
\period
\eqe
In terms of
$a := \sn^2A_1,b := \sn^2A_2$, this 
is equivalent to
\eqb
(a+b-ab)(a+b-k^2ab)(a+b-(1+k^2)ab)=0 \comma
\eqe
the solutions of which are 
\eqb
\mbox{(i)}\ \cn\,A_1\,\cn\,A_2=\pm 1,\quad
\mbox{(ii)}\ \dn\,A_1\,\dn\,A_2=\pm 1,\quad
\mbox{(iii)}\ {\rm cd}\,A_1\,{\rm cd}\,A_2=\pm 1. \label{Virasoro}
\eqe
In each of these three cases, one finds that
\eqb
\mbox{(i)}\ u_0=-2k^2,\quad
\mbox{(ii)}\ u_0=-2,\quad
\mbox{(iii)}\ u_0=0, \label{u0}
\eqe
and 
\eqb
\mbox{(i)}\ u=2k^2U^+U^-\cn^2 X,\quad
\mbox{(ii)}\ u=2U^+U^-\dn^2 X,\quad
\mbox{(iii)}\ u=-2k^2U^+U^-\sn^2 X.
\eqe

The final condition to be imposed is the reality condition, 
for which we need to know the allowed identifications between 
$ \varphi $ and $Y$.
In order to analyze these, 
we note that, from $\det \varphi = \det Y = 1$ and the equations of motion, 
the two matrices should be related by constant $ SL(2,\bbC)$ matrices $U,V $ as
$U \varphi V = Y$. This implies that $Y^{-1} dY = V^{-1} \varphi^{-1} d\varphi V$,
and that the tangent spaces of $Y$ and $\varphi$ are isomorphic. Since
$Y$ is an $SL(2,\bbR)$ matrix,  $\varphi$ should generically  be an element of
 $SL(2,\bbR)$ or $SU(1,1)$. Therefore, 
there are two cases for the reality condition: 
\eqb
     && ({\rm I}) \quad  \ \varphi_{j}^* = \varphi_{j} \comma \  (\phiinv_{j})^* = \phiinv_{j}
       \qquad \  {\rm for } \quad \varphi \in SL(2,\bbR) \comma \nn \\ 
     && ({\rm II}) \quad  \varphi_{1}^* = \phiinv_{1} \comma \quad \varphi_{2}^* = - \phiinv_{2}
     \qquad {\rm for } \quad \varphi \in SU(1,1)  \comma \label{realcond}
\eqe
(up to the exchange of $\varphi_{1}, \phiinv_{1}$ and $\varphi_{2}, \phiinv_{2}$).
In each case, the AdS solution $Y$ is identified with $\varphi$ as
\eqb
     ({\rm I})   \quad  \varphi = Y \comma \qquad
     ({\rm II})   \quad  \varphi = M^{-1} Y M
     = \biggl(\begin{array}{cc}Y_{-1} + iY_{0} & Y_{1}+iY_{2} \\Y_{1}-iY_{2} & 
     Y_{-1}-iY_{0}\end{array}\biggr)\comma 
\eqe
up to $SL(2,\bbR)$ and $SU(1,1)$ transformations, respectively, where
$ M =  \frac{1}{\sqrt{2}} \biggl(\begin{array}{cc} 1& i \\i & 1\end{array}\biggr)$.
In both cases, the potential takes the form $u = -\del_{+}\vec{Y} \cdot \del_{-}\vec{Y}$,
from which one can read off the conformal factor of the induced metric and hence
the curvature of the surface described by the solution.

In the following, we set $q = e^{\pi i \tau}$ 
($ \tau = iK'/K$) 
in the theta functions to be real, which implies $ 0 \leq k^{2} \leq 1$.
When $q$ is complex, the reality conditions may not be satisfied.

\msection{Solving reality and Virasoro conditions}

In this section, we solve the reality and  Virasoro conditions which are
listed in the previous section. First, we concentrate on the case where 
$\varphi \in SL(2,\bbR)$. The other case with $\varphi \in SU(1,1)$ is
discussed later. 

\msubsection{reality condition}
 When $\varphi \in SL(2,\bbR)$, the reality condition is (I) in (\ref{realcond}). 
For this to be satisfied for arbitrary
$\sigma_{\pm}$, the theta functions $\varth_{0}(X), \varth_{0}(X+A_{j}) $ should be
  real or purely imaginary (after extracting the exponential factors
   due to possible shifts
  in $(X,A_{j})$ by $iK')$. 
  This implies that $(X, A_{j})$ are real or purely imaginary up to 
  the shifts $n K + i m K'$, under which $\varth_{a}$ transform 
  as $\varth_{a}(u + n K + i m K') = $ (factor)$\times \varth_{b}(u)$ 
  according to (\ref{thetashift}).
  Furthermore, since $\varth_{a}(u+2K) = \pm \varth_{a}(u)$, $\varth_{a}(u +2 i K') 
  = \pm e^{-\pi i (\frac{u}{K}+ \tau)} \varth_{a}(u)$, we have only to consider 
  $0, K, iK', K + iK'$ as the shifts: other cases reduce to these cases by 
  absorbing the factors into $r_{j}$ and $p^{\pm}_{j}$. Consequently, 
  it is enough to assume that $A_{j}$  are in the fundamental region spanned by
  $(0,2K,2iK', 2K+2iK')$ with  segments $[2K, 2K+2iK'], [2iK', 2K + 2iK']$ removed.

Therefore, we have four cases of $(X,A_j)$:
\eqb
 &&  (1) \quad X \in \bbR \  \ {\rm or}  \  \ \bbR + iK'  
   \quad {\rm and} \quad  A_j=a_j \ (a_j \in \bbR) \comma \nn \\
  && (2) \quad X \in \bbR \  \ {\rm or}  \  \ \bbR + iK' 
   \quad {\rm and} \quad  A_j=a_j +iK' \ (a_j \in \bbR) \comma \nn \\
  && (3) \quad X \in i\bbR \  \ {\rm or}  \  \ i\bbR + K 
  \quad {\rm and} \quad  A_j= ia_j \ (a_j \in \bbR) \comma  \label{XA} \\
  && (4) \quad X \in i\bbR  \  \ {\rm or}  \  \ i\bbR + K 
  \quad {\rm and} \quad  A_j=ia_j +K \ (a_j \in \bbR) \period \nn 
\eqe
In addition, after the possible shifts of $iK'$ in $(X,A_j)$ are taken into account,
real solutions for $\varphi \in SL(2,\bbR)$ must be transformed into the canonical
form where $r_j,\rinv_j$ and exponentials in $\varphi$ are real.
These impose restrictions on $\beta^\pm_{j}$. 

Let us discuss these conditions in more detail,
e.g., in case (2). In this case, $U^+ \sigma_+ + U^- \sigma_-$ is real, which implies
$(U^{+})^{*} = U^{-}$ and $X_{0} -K \in \bbR$ or $\bbR + iK'$.
When $X_0 -K \in \bbR + iK'$, the shift of $iK'$ results in a constant factor to the ratio
of the theta functions, which
we absorb into $r_j, \rinv_j$.
As for the shift $iK'$ in $A_j$, extracting it from $\varth_0$ gives
\eqb
  &&\varphi_j \sim \frac{\theta_{1}(X+a_j)}{\theta_{0}(X)} \, e^{q^{+}_j\sigma_+ + q^{-}_j \sigma_-}
  \comma \ \quad q^{\pm}_j = p^{\pm}_j - \frac{\pi  i}{2K} U^{\pm} \comma \nn \\
  && Z(a_j + iK')  
  = -\frac{\pi i}{2K} + Z_{1}(a_j) \comma \quad Z_1(z) :=\del_z \ln \varth_1(z) \comma   
\eqe
and similarly for $\phiinv_j$ with the signs of $a_j, q_j^\pm$ flipped.
The exponent after the shift should be real and thus $(q_j^{+})^{*} = q_j^{-}$.
Note that $(q_j^{+}/U^{+})^{*} = q_j^{-}/U^{-}$, $Z_{1}(a_j) \in \bbR$, and the conditions
from the equations of motion (\ref{betaeom}) read
\eqb
   \beta_j^{+} + \beta_j^{-} = 2 \frac{\dn\, a_j \, \cn\, a_j}{\sn\, a_j} \in \bbR \comma
   \quad 
   \beta_j^{+} \beta_j^{-} = \ns^{2}a_j + u_{0} \in \bbR . \label{betapm1}
\eqe
On the other hand, from the definition of $\beta_j^\pm$, (\ref{betadef}), it follows that 
\eqb
   \beta_j^{\pm} =  Z_{1}(a_j) +\frac{q_j^{\pm}}{U^{\pm}}  \comma \label{betapm2}
\eqe
and hence $(\beta_j^+)^*= \beta_j^-$.
For given $k, a_j,U^\pm$, the real part of $\beta_j^{\pm}$ (or $p_j^{\pm}/U^{\pm}$) 
is determined by the  first equation in (\ref{betapm1}), 
whereas the imaginary part is consistently determined by the second, if
\eqb
   \Delta_{\beta_j}:=\frac{1}{4}(\beta_j^{+} - \beta_j^{-})^{2} = \frac{1}{\sn^{2} A_j} -(1+k^{2} + u_{0})
   \leq 0 \comma
\eqe
with $1/\sn^2 A_j = k^2 \sn^2 a_j$.

Similarly analyzing other cases, we find that the reality condition imposes
\eqb
   \Delta_{\beta_j} \leq 0 \quad {\rm for \quad (1) (2)} \comma  \qquad 
    \Delta_{\beta_j} \geq 0 \quad {\rm for \quad (3) (4)} \period
\eqe
Applying the value of $u_{0}$ in (\ref{u0}),
 these are solved in each case, which imposes the following 
 conditions: 
\begin{namelist}{cccccc}
\item[\quad (1)] $A_j = a_j \ (a_{j} \in \bbR)$ 
 \vspace*{0.5ex} \\
(i)  $k' = \sn^{2}a_j = 1$; \ (ii)  (no solutions); \ (iii) $\sn^{2}a_j \geq \frac{1}{1+k^{2}}$. 
\item[\quad (2)]  $A_j = a_j  +iK'\ (a_j \in \bbR)$ 
 \vspace*{0.5ex} \\
(i)  $\sn^{2} a_j \leq \frac{(k')^{2}}{k^{2}}$; \ (ii)  (no solutions); \ (iii) (automatic). 
\item[\quad (3)]  $A_j = ia_j\ (a_j \in \bbR)$
 \vspace*{0.5ex} \\
(i)  (no solutions); \ (ii)  $\sn^{2}(a_j,k') \geq \frac{1}{1+(k')^{2}}$; \ (iii) (no solutions). 
\item[\quad (4)]  $A_j = ia_j +K \ (a_j \in \bbR)$
 \vspace*{0.5ex} \\
(i)  $\sn^{2}(a_j,k') \leq \frac{k^{2}}{(k')^{2}}$; \ (ii) (automatic) ; \ (iii) $ k= a_j =0$. 
\end{namelist}
In the table, ``no solutions'' indicates the cases  where the solutions do not  exit, 
whereas ``automatic"  indicates the cases where the reality condition is automatically 
satisfied  and  imposes no
restrictions.  The cases where $\beta_j^\pm$ are diverging have also been excluded.
We have also omitted the values  of $X$ in the above.

We remark that, when considering both $\varphi_1,\phiinv_1$ and $\varphi_2,\phiinv_2$,
$X$ is common and only the combinations among cases (1) and (2), or (3) and (4)
are allowed.

\msubsection{Virasoro condition}

{}From the discussion in the previous section,
we find that there are six cases of the combinations of $(A_{1},A_{2})$.
In each combination, there are three possibilities of satisfying the Virasoro
condition as in (\ref{Virasoro}). It is straightforward to write down 
the explicit from of the condition in each case and check whether
it has solutions or not.

{}For example, when $A_{1}= ia_{1}, A_{2} = ia_{2} $ ($a_{1}, a_{2} \in \bbR$), 
the condition of case (i) in (\ref{Virasoro}) reads $ \nc(a_{1},k') \, \nc(a_{2},k') = \pm 1$.
Since $\nc^{2} u \geq 1$ for real $u$, the condition is satisfied only when
$a_{1} = a_{2} = 0$. When $A_{1,2} = a_{1,2} + iK'$ ($a_{1,2} \in \bbR$),
the condition of case (i) in (\ref{Virasoro}) reads $-k^{-2}\ds\, a_{1} \, \ds\, a_{2} =\pm 1$.
Since $\ds^{2} u \geq (k')^{2}$ for real $u$, the condition has solutions
when $ 1/2 \leq k^{2}$.

Repeating similar analysis for all cases, one finds that the Virasoro constraints 
 impose the following conditions:
\begin{namelist}{cccccc}
\item[\quad 1-1.] $A_{1} = a_{1}, \ A_{2} = a_{2}$  \ $(a_{j} \in \bbR)$ 
 \vspace*{0.5ex} \\
(i) $ a_{1,2} = 0 $; \ (ii)   $a_{1,2} = 0 $ or $k=0$; \ (iii) $ a_{1,2} = 0 $ or $k=1$. 
\item[\quad 2-2.] $A_{1} = a_{1} + iK', \ A_{2} = a_{2} + iK' $  \ $(a_{j} \in \bbR)$ 
  \vspace*{0.5ex}\\
(i) $k^2 \sd\, a_{1} \, \sd\, a_{2} =\pm 1$ and $ \half  \leq k^{2}$; 
 \ (ii)   $\sc\, a_{1} \, \sc \, a_{2} = \pm 1$; \ (iii)  $k=1$.
\item[\quad 1-2.] $A_{1} = a_{1}, \ A_{2} = a_{2} + iK' $  \ $(a_{j} \in \bbR)$ 
  \vspace*{0.5ex} \\
  (i)  (no solutions); 
   \ (ii)  (no solutions); \  (iii)  $ k \, \dc \, a_{1} \, \cd \, a_{2} = \pm 1$.
\item[\quad 3-3.] $A_{1} = ia_{1}, \ A_{2} = ia_{2}  $  \ $(a_{j} \in \bbR)$ 
  \vspace*{0.5ex} \\
  (i)  $a_{1,2}  = 0$; 
   \ (ii)  $ a_{1,2}  = 0 $ or $k=0$; \ (iii)  $ a_{1,2} = 0 $ or $k=1$.
\item[\quad  4-4.] $A_{1} = ia_{1}+K, \ A_{2} = ia_{2}+K$  \ $(a_{j} \in \bbR)$ 
  \vspace*{0.5ex} \\
  (i)  $ (k')^{2}\sd(a_{1},k') \, \sd(a_{2},k') =\pm 1$ and $ \half  \geq k^{2}$; 
   \ (ii)  $k=0$; \  (iii)  $\sc(a_{1},k') \, \sc(a_{2},k') = \pm 1$.
\item[\quad 3-4.] $A_{1} = ia_{1}, \ A_{2} = ia_{2}+K$  \ $(a_{j} \in \bbR)$ 
  \vspace*{0.5ex} \\
  (i)  (no solutions); 
   \ (ii) $ k' \, \dc(a_{1},k') \, \cd(a_{2},k') = \pm 1$; \ (iii)  (no solutions).  
\end{namelist}
\msubsection{classification}

Combining the tables in the previous two subsections, we can determine
the allowed cases and their conditions:
\begin{namelist}{cccccc}
\item[\quad 1-1.] $A_{1} = a_{1}, \ A_{2} = a_{2}$  \ $(a_{j} \in \bbR)$ 
 \vspace*{0.5ex} \\
(i) (no solutions); \ (ii)   (no solutions); \ (iii) $k=1$ and $\sn^{2}a_{1,2} \geq \half$. 
\item[\quad 2-2.] $A_{1} = a_{1} + iK', \ A_{2} = a_{2} + iK' $  \ $(a_{j} \in \bbR)$ 
  \vspace*{0.5ex}\\
(i)  $k^{2}= \half$ and $\sn^{2}a_{1,2} = 1$; 
 \ (ii)   (no solutions); \  (iii) $k=1$.
\item[\quad 1-2.] $A_{1} = a_{1}, \ A_{2} = a_{2} + iK' $  \ $(a_{j} \in \bbR)$ 
  \vspace*{0.5ex} \\
  (i)  (no solutions); 
   \ (ii)  (no solutions); \ (iii)  $ k \, \dc \, a_{1} \, \cd \, a_{2} = \pm 1$ and 
   $\sn^{2}a_{1} \geq \frac{1}{1+k^{2}}$.
\item[\quad 3-3.] $A_{1} = ia_{1}, \ A_{2} = ia_{2}  $  \ $(a_{j} \in \bbR)$ 
  \vspace*{0.5ex} \\
  (i)   (no solutions); 
   \ (ii)  $k=0$ and $\sn^{2}(a_{1,2},k') \geq \half$; \  (iii)   (no solutions).
\item[\quad 4-4.] $A_{1} = ia_{1}+K, \ A_{2} = ia_{2}+K$  \ $(a_{j} \in \bbR)$ 
  \vspace*{0.5ex} \\
  (i)  $k^{2} = \half$ and $\sn^{2}(a_{1,2},k') = 1$; 
   \ (ii)  $k=0$; \ (iii)  (no solutions).
\item[\quad 3-4.] $A_{1} = ia_{1}, \ A_{2} = ia_{2}+K$  \ $(a_{j} \in \bbR)$ 
  \vspace*{0.5ex} \\
  (i)  (no solutions); 
   \  (ii) $ k' \, \dc(a_{1},k') \, \cd(a_{2},k') = \pm 1$ and 
   $\sn^{2}(a_{1},k') \geq \frac{1}{1+(k')^{2}}$; \\
    (iii)  (no solutions).  
\end{namelist}

We note that the result is symmetric between the real and the imaginary $A_j$.
This is a consequence of the modular transformation $\tau \to -1/\tau$ with purely 
imaginary $\tau$. In fact, one can check that $\varphi$ in the first three cases
are mapped to the last three, up to certain  factors which can be absorbed 
into the exponential factors  and  the normalization constants of $\varphi$. 
Some asymmetries 
in the intermediate stage of the analysis are  due to having started with
the fixed exponents in $\varphi$. 

From this result,  one finds that the allowed solutions fall into three types. One  is 
the solution with $k \neq 0,1$ and $A_1 \neq A_2$ as in 1-2 (iii) and
3-4 (ii). Such solutions are expressed by  the elliptic  functions, as
we initially intended.  We call this type of solutions ``elliptic solution''.

The second one is the  solution with $k=0$ or $1$
and $A_1 \neq A_2 $. In this case, the elliptic functions degenerate
and the solutions become simpler. One has to be a little careful 
in taking $k=0,1$, since $K',K$ are singular, respectively. 
For $k=0$ as in 3-3 (ii) and 4-4 (ii), 
$q = e^{i \pi \tau}$ is vanishing and  $\varth_0(X)$ reduces to a constant for finite $X$.
However, if we take $k\to 0$ after  shifting  $X$, which is 
imaginary in these cases, by $iK'(\to i\infty)$, the ratio of $\varth_0$'s becomes a ratio 
of the hyperbolic functions. 
For $k=1$ with $q \to 1$ as in 1-1 (iii) and 2-2 (iii),
by making use of  the modular transformation
 $\tau \to -1/\tau $, one finds 
that for finite $X$ the ratio of $\varth_0$'s becmes a ratio of  the hyperbolic functions.
 However, if we take $k\to 1$ after shifting $X$ by $K (\to \infty)$, which is allowed 
 in these cases, the ratio of $\varth_0$'s becomes a constant. 
Thus, depending on the way to take the limit, the degenerate solutions 
reduce to  (a) the known solutions with $\varphi_j,\phiinv_j \sim$ const.$\times $(exponentials),
or (b) the solutions with $\varphi_j,\phiinv_j \sim$ 
(ratio of hyperbolic functions)$\times $(exponentials). 
In the latter case, the potential $u$ is not constant, and the 
minimal surface spanned by $Y$  is not flat. 
We call the former type ``exponential solution'', and the latter 
``hyperbolic solution''.

The third type is the solution with $A_1 =  A_2$, which we
have not considered so far, since the normalization condition (\ref{rr})
becomes singular.  In this case,  (\ref{phiphi}) implies
that the  only possibilities to satisfy
the normalization condition of $\varphi$ is $k=0,1$ or $A_{1,2} =0$, since
the $X$-dependence has to be canceled. 
Thus,  cases 2-2 (i), 4-4 (i) are excluded, though
we left them in the table taking into account a possibility that 
they  could be regarded as limiting cases.
When $A_{1,2}= 0$, the solutions reduce to the exponential type.
When $k=0,1$, as discussed above, the solutions become of the exponential
or the hyperbolic/trigonometric type. In the latter case,
 it turns out that one has to set $A_1 = A_2 =0$
 to satisfy the normalization 
condition. In sum, if $A_1=A_2$, 
only the solutions
of the exponential type are allowed.

For $k=0,1$ or $A_{1} = A_{2}$, one may take appropriate limits from the generic
cases to write down the solutions. However, it is more straightforward to 
start with the generic form
of the solutions in these cases, and determine them as in section 2.

\newpage
\msubsection{$SU(1,1)$ case}

So far, we have considered the case where $\varphi \in SL(2,\bbR)$. As discussed in
section 2, there is another case with $\varphi \in SU(1,1)$. 
The reality and Virasoro conditions are analyzed similarly.

First, from the reality condition (II) in (\ref{realcond}), one finds that there are
four cases of $(X,A_j)$: 
\eqb
 &&  (1') \quad X \in i \bbR \  \ {\rm or}  \  \ i \bbR + K  
  \quad {\rm and} \quad  A_j=a_j \ (a_j \in \bbR) \comma \nn \\
  && (2') \quad X \in i \bbR \  \ {\rm or}  \  \ i \bbR + K 
  \quad {\rm and} \quad  A_j=a_j +iK' \ (a_j \in \bbR) \comma \nn \\
 && (3') \quad X \in \bbR \  \ {\rm or}  \  \ \bbR + iK'  
   \quad {\rm and} \quad  A_j= ia_j \ (a_j \in \bbR) \comma  \\
  && (4') \quad X \in \bbR  \  \ {\rm or}  \  \ \bbR + iK'  
   \quad {\rm and} \quad  A_j=ia_j +K \ (a_j \in \bbR) \comma \nn 
\eqe
In addition, the normalization constants should satisfy 
$ r_{j}^{*} = \pm \rinv_{j}$ and the
exponentials in $\varphi_{j}$ and $\phiinv_{j}$ should be complex conjugate to 
each other (after the possible shifts of $iK'$ in  $(X,A_{j})$).

In any of these cases, the combinations  of $p_j^{\pm}$ and $U^{\pm}$ 
satisfy the same relations as the corresponding ones in the  $SL(2,\bbR)$ case.
For example, in case  $(1')$, $(p_j^{+}/U^{+})^{*} = p_j^{-}/U^{-}$, though 
$p_j^\pm, U^\pm$ have different relations  $(p_j^{+})^{*} = -p_j^{-}$
and $(U^{+})^{*} = -U^{-}$.
Thus, the constraints from the reality condition are the same.

The Virasoro condition is irrelevant of which embedding we use,  $SL(2,\bbR)$ or $SU(1,1)$.
Therefore, the allowed cases are read off from the same table as in the 
$SL(2,\bbR)$ case in section 3.3. In the $SU(1,1)$ case, the condition on $r_{j},\rinv_{j}$ implies
\eqb
  r_{1}\rinv_{1} r_{2} \rinv_{2} \leq 0 \period
\eqe
This may give further constraints on the parameters, e.g., on $X_{0}$.
When $r_{1}\rinv_{1} < 0$ and $r_{2}\rinv_{2} > 0$, we need to exchange
$\varphi_{1}, \phiinv_{1}$ and $\varphi_{2},\phiinv_{2}$. 
For the Euclidean world-sheet, which results in space-like
surfaces, one finds no solutions eventually.

\msection{Examples of solutions}
Our main motivation to studying the AdS  string solutions
is the application to the scattering amplitudes 
 in the super Yang-Mills theory.
With this in mind, we discuss the obtained solutions.

\msubsection{searching for cusp solutions}

Before going into details, let us summarize some general
properties of the solutions in relation to the cusp solutions with null boundaries. 
First, when  $\varphi \in SU(1,1)$, the exponential part of $\varphi$
is complex and, since, e.g., $Y_{-1} =$ Re$\, \varphi_1$, 
the solutions are generally 
rapidly oscillating near the world-sheet boundary $|\sigma_\pm| \gg 1 $. 
Thus, to search for the cusp
solutions, one should look into the case with  $\varphi \in SL(2,\bbR)$
(unless the world-sheet is consistently restricted). This case also includes oscillating
solutions. For example, in case (2)  with real $X$ in (\ref{XA}), the solution has a factor
$\varth_2(X+a)$
and this is oscillating. In case (1) with real $X$, the solution
has an oscillating factor $\varth_0(X+a)$ but, since this does not change the sign,
the oscillation is harmless (as can be checked by the modular transformation). 
For imaginary $X$, if the solutions contain the factors of
$\varth_{0,1}$, they oscillate, whereas if the factors are $\varth_{2,3}$, they 
do not. The limiting cases with $k=0,1$ are similarly considered.

Once one finds the solutions which grow large without harmful
oscillation near the world-sheet boundary,
they are good candidates of the cusp solutions. Though some of the cusps
are generally at the infinity in the boundary Poincar\'e coordinates, 
$x_{\pm} = (Y_{1}\pm Y_{0})/(Y_{-1}+Y_{2})$, 
they can be  brought to  finite
points by an $SL(2,\bbR)$ transformation:
\eqb
  Y = U \varphi V \comma \quad U = \Biggl(\begin{array}{cc}a & b \\c & d\end{array}\Biggr)
  \comma \  V = \Biggl(\begin{array}{cc} \delta  & -\beta \\ -\gamma & \alpha\end{array}\Biggr)
  \comma
\eqe
with $\det U = \det V = 1$.
To see this, we trace a contour with a large radius in the world-sheet.
Supposed that one of $\varphi_{j}, \phiinv_{j}$ alternatively 
becomes dominant along the contour, one finds that the contour is mapped  
to a rectangular in the $ x_{\pm}$-plane which has null boundaries and 
four cusps at  $(x_{+},x_{-}) 
= (\frac{c}{a}, -\frac{\beta}{\delta}),
(\frac{c}{a}, -\frac{\alpha}{\gamma}),(\frac{d}{b}, -\frac{\alpha}{\gamma}),(\frac{d}{b},
 -\frac{\beta}{\delta})$.
If $\varphi$ shows a more intricate behavior, more cusps and null boundaries 
may appear.
We note that one should choose  the  $SL(2,\bbR)$ transformation so that
the Poincar\'e radial coordinate $1/(Y_{-1}+Y_{2})$ is non-negative: otherwise,
the interpretation of the solution in the Poincar\'e corrdinates may be subtle.

\msubsection{elliptic solutions}

{}From the discussion in the above, we find that the elliptic solutions
in 1-2 (iii) and 3-4 (ii) are harmfully oscillating solutions. However, in the limit $k \to 1$
for 1-2 (iii) and $k \to 0$ for 3-4 (ii), 
they become the exponential or the hyperbolic solutions in which  the oscillation 
disappears. This is because, e.g., for 1-2 (iii), the period of the oscillation
is of order $K$, and this is diverging as $k \to 1$; only the strip with the width
of order $K$ survives after taking the limit. Conversely, this shows that unwanted oscillation
might be eliminated by restricting the world-sheet in some region.
In our case, simply taking the world-sheet to be the strip does not give
cusp solutions with null boundaries, since the two sides of the strip are mapped
to the boundary of the surface in $AdS$ which is  not null.
However, this may be a useful tip to further search 
for the cusp solutions.  These elliptic solutions 
are regarded as elliptic generalizations of the known exponential solutions.

\newpage
\msubsection{degenerate solutions}
The degenerate solutions of the hyperbolic/trigonometic type with $k=0,1$ 
are new solutions. 
As mentioned at the end of section 3.3, 
instead of taking appropriate limits from the generic cases, 
one  may  start with the generic form of the solutions in this case,
\eqb\label{degensol}
\varphi_j\Eqn{=}
  r_j\frac{\cosh(\mu^+\sigma_+ + \mu^-\sigma_- + \alpha_j)}
          {\cosh(\mu^+\sigma_+ + \mu^-\sigma_-)}
          e^{p_{j}^{+} \sigma_{+} + p_{j}^{-} \sigma_{-}}, \nn \\
\phiinv_j \Eqn{=}
  \rinv_j\frac{\cosh(\mu^+\sigma_+ + \mu^-\sigma_- - \alpha_j)}
          {\cosh(\mu^+\sigma_+ + \mu^-\sigma_-)}
          e^{-(p_{j}^{+} \sigma_{+} + p_{j}^{-} \sigma_{-})},
\eqe
and determine them as in section 2.
Indeed, one finds that the normalization condition and the equations of motion
give
\eqb
  1 = r_{1} \rinv_{1} + r_{2} \rinv_{2} \comma \quad 
  0 = r_{1}\rinv_{1} \sinh^{2}\alpha_{1} + r_{2}\rinv_{2} \sinh^{2}\alpha_{2} 
  \comma \label{degnorm}
\eqe
and
\eqb
   p_{1}^{+}  p_{1}^{-} =  p_{2}^{+}  p_{2}^{-} \comma \quad
    0= (p_{j}^{+} \mu^- + p_{j}^{-} \mu^+) \tanh \alpha_{j} + 2 \mu^+ \mu^- \quad  (j=1,2)
  \comma \label{degeom}
\eqe
respectively, 
whereas the Virasoro condition imposes
\eqb
 && 0 =  (p_{1}^{\pm})^{2} r_{1} \rinv_1 +  (p_{2}^{\pm})^{2} r_{2} \rinv_2  \comma \nn \\
 && 0 = (p_{1}^{\pm})^{2} - (p_{2}^{\pm})^{2} 
  + 2\mu^{\pm} \Bigl( \frac{p_{1}^{\pm}}{\tanh \alpha_{1}} 
  - \frac{p_{2}^{\pm}}{\tanh \alpha_{2}} \Bigr) \period \label{degvirasoro}
\eqe

Since we are interested in the cusp solutions  with real $\varphi$, we impose 
the reality condition $(p_j^\pm)^* = p_j^\mp$, $(\mu^\pm)^* = \mu^\mp$.
It turns out  that the  solutions to the constraints (\ref{degnorm})-(\ref{degvirasoro}) 
are essentially unique (up to conformal transformations of $\sigma_\pm$ etc.),
and given by
\eqb
 && \mu^\pm = \frac{1}{\sqrt{2}}\,  e^{\pm i\theta} \comma \quad 
  p^\pm_1 = 1\comma \quad  \ p^\pm_2 = \mp i \comma \quad 
  r_1 \rinv_1  =   r_2 \rinv_2 = \frac{1}{2} \comma \nn  \\
 &&  \tanh\alpha_1  
 =-\frac{1}{\sqrt{2}\cos\theta} \comma \quad 
\tanh\alpha_2 
 =\frac{1}{\sqrt{2}\sin\theta} \period
 \eqe
These give a class of real and non-oscillating solutions of the form,\footnote{
Shifting the argument of $\cosh$ in (\ref{degensol}) by $\pi i/2$ gives
solutions with $\coth B$.
}
\eqb
\varphi_1\Eqn{=}\frac{1}{\sqrt{\cos 2\theta}}
\left(\cos\theta - \frac{1}{\sqrt{2}}\tanh B\right) e^{t}, \nn \\
\varphi_1^\sigma\Eqn{=}\frac{1}{\sqrt{\cos 2\theta}}
\left(\cos\theta + \frac{1}{\sqrt{2}}\tanh B\right) e^{-t}, \nn \\
\varphi_2\Eqn{=}\frac{1}{\sqrt{\cos 2\theta}}
\left(\sin\theta + \frac{1}{\sqrt{2}}\tanh B\right) e^{s}, \label{degeneratesol}\\
\varphi_2^\sigma\Eqn{=}\frac{-1}{\sqrt{\cos 2\theta}}
\left(\sin\theta - \frac{1}{\sqrt{2}}\tanh B\right) e^{-s}, \nn
\eqe
where
$ B  =\frac{\cos\theta}{\sqrt{2}} t - \frac{\sin\theta}{\sqrt{2}}s $,
$ \sigma_{\pm}=(t \pm i s)/2 $, and  we have assumed $\cos 2\theta > 0$.
(The case with $\cos 2 \theta < 0$ is similar.)
The potential $u$ reads 
\eqb
 u = - \tanh^2 B \period \label{degenpot}
\eqe

\begin{figure}[t]
         		\begin{center}
		         \includegraphics[width=7cm]{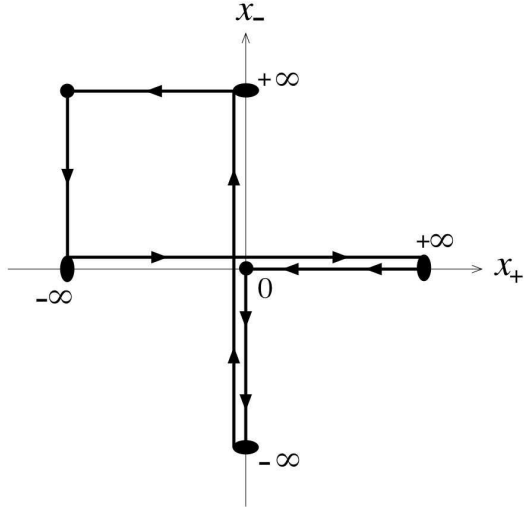}
		      \caption{ Boundary of minimal surface described by (\ref{degeneratesol})
		         in $(x_{+}, x_{-})$-plane.  A contour with a large radius 
		         in the world-sheet is mapped to the  $(x_{+}, x_{-})$-plane
		         along the arrows. 
		         }
	         \end{center}
	         \vspace*{-0ex}
\end{figure}

This class of solutions includes solutions which have 
four cusps, two horns and six null boundaries, 
 among which two pairs are collinear.
We remark that  all the boundaries are null.
To check these properties, 
we first restrict to the case  where $\cos \theta > 1/\sqrt{2}$ so that
the Poincar\'e radial coordinate $r = 1/\varphi_1 $ is non-negative.
 (In the other case with $\cos \theta < -1/\sqrt{2}$, we have only to
flip the signs of $r_j, \rinv_j$.)
Next, we note that the AdS boundary is given by $|Y_{-1} + iY_0 | \to \infty$.
Plugging the solution (\ref{degeneratesol}) into $|Y_{-1} + iY_0 | $, 
we then find that the world-sheet boundary
where  $| t | $ or $ | s |  \to \infty $ is mapped to the AdS boundary
unless $ \cos \theta, \sin \theta = 0, \pm \frac{1}{\sqrt{2}} $. When $\theta$ takes 
such a generic value, similarly to the discussion in section 4.1 we find that 
the image of the world-sheet boundary traces six null segments 
in the $(x_+,x_-)$-plane. 
Concretely, the contour $t = \rho \sin \omega, \, s = \rho \cos \omega$
with $\rho \to \infty$ is mapped to $(-\infty,0) \to (\infty, 0) \to (0,0) \to (0,-\infty)
\to (0,\infty) \to (-\infty, \infty)  \to (-\infty, 0)$ as $\omega$ varies from $0 $ to $2\pi$.
The resultant 
boundary of the surface is not convex, but 
crossed and folded as in Fig.1.
Among the six end-points of the segments, 
$(0,0), (0, -\infty), (-\infty, \infty), (-\infty, 0)$ are the cusps and  
$(\infty, 0),(0, \infty)$ are the tips of the two horns.
The essence  in producing the six null boundaries is that the change of the sign 
of $\tanh B$ ``splits'' the cusps, which, in the $x_\pm$-plane, 
is observed  as  the transitions $(-\infty,0) \to (+\infty, 0)$ and $(0,-\infty)  \to (0,+\infty)$.
In these transitions, the surface boundary has to keep touching the AdS
boundary. Since the solution has two pairs of collinear null boundaries,
one expects that it gives the scattering amplitudes at strong coupling in a collinear limit.

When $ \sin \theta =0$, the surface boundaries 
mapped from $ t = 0$
do not reach the AdS boundary, and they  form two  boundaries inside $AdS_{3}$.
Consequently, the solution describes a surface which has
four null boundaries at the AdS boundary, and two  boundaries inside AdS.
The surface pinches at a point where these
two boundaries intersect each other.
The shape of the surface is obtained by diagonally cutting a  four-cusp surface
and twisting it.

{}From the potential in (\ref{degenpot}), one finds that
the surface has non-trivial curvature. This shows a clear difference 
from the four-cusp solution in \cite{Kruczenski:2002fb,Alday:2007hr},
where the potential is constant and hence the corresponding surface is flat.
In fact, the two solutions are not related to each other 
by simple transformations:
First, they cannot be related by an $SO(2,2)$ transformation, 
since the potential $u = -\del_{+}\vec{Y}\cdot \del_{-}\vec{Y}$ is invariant.
Second, as long as we work with the Euclidean world-sheet, 
the allowed world-sheet analytic continuation is the continuation 
of both $t$ and $s$, which results in $\del_{\pm} \to \pm i \del_{\pm}$. 
Thus, $u$ is invariant up to a sign and renaming the world-sheet coordinates.
Third, one may generate a new solution by a target-space analytic continuation
such as $Y_{a} \to i Y_{a}$ together with the world-sheet analytic continuation
as in \cite{Kruczenski:2007cy}.
However, the potential should again be invariant (up to a sign and renaming of 
the world-sheet coordinates) in order to keep
the equations of motion invariant. Finally, if the potential $u$ has a factorized
form $f(\sigma_{+})g(\sigma_{-})$, it may be brought to a constant by 
a world-sheet conformal transformation, but this is not possible 
for the degenerate solution. 

\msection{\mathversion{bold} Adding $S^{1}$: elliptic four-cusp solutions}
The analysis so far can be generalized to 
the case of the strings in $AdS_{5} \times S^{5}$. Here, for simplicity,
we consider the case of $AdS_{3} \times S^{1}$.

In an appropriate gauge, the $S^{1}$ field is set to be $W = \kappa_{+} \sigma_{+} +
\kappa_{-} \sigma_{-} $. The reality of $W$ requires $(\kappa_{+})^{*} = \kappa_{-}$.
Adding $S^{1}$ does not change the reality condition on $Y$, but that changes the
Virasoro constraints  to
\eqb
   \sum_{j=1}^2\del_{\pm} \varphi_j\del_{\pm}\varphi^\sigma_j  = \kappa_{\pm}^{2} 
   \period
\eqe
Similarly to the case without $S^{1}$, linear combinations of these  give
\eqb
  && u_{0} = 2\left(\frac{1}{\sn^2 A_1}+\frac{1}{\sn^2 A_2}-1-k^2\right) - 
  \Bigl(  \frac{U^{-}}{U^{+}} \kappa_{+}^{2} +  \frac{U^{+}}{U^{-}} \kappa_{-}^{2}\Bigr) \comma
   \\
 &&  2U^+U^-
\frac{\sn^2 A_1\,\sn^2 A_2}{\sn^2 A_2-\sn^2 A_1}
\sum_{j=1}^2(-)^{j+1}\frac{\cn\,A_j\,\dn\,A_j}{\sn^3 A_j} (\beta^{+}_{j} - \beta^{-}_{j})
= \Bigl(  \frac{U^{-}}{U^{+}} \kappa_{+}^{2} -  \frac{U^{+}}{U^{-}} \kappa_{-}^{2}\Bigr)
\period \nn
\eqe

Because of the change of the Virasoro constraints, 
the allowed solutions for $Y$ also change.
Though they can be classified as in section 3, we do not go into details.
However, we know that, in order to find cusp solutions, we have only to look into 
the cases without harmful oscillation. Among the elliptic cases, they are 
1-1 or 3-3 in section 3.2, which 
are related to each other by 
the modular transformation. In the following, we take 3-3.
It turns out that this case indeed gives four-cusp solutions with null boundaries 
which are expressed by
the elliptic  functions.

For example, for $k = 0.7, U^\pm = i, A_{1} = i K'/2, \kappa_{\pm} = \sqrt{12/5}(1\pm i)$, 
the Virasoro condition gives $A_{2} = 1.277...$ and $u_{0}=-4.781... \,$. 
Further setting $X_0 = 0$, the theta function takes the form
$\varth_0(X +ia)= \varth_3(i t+ ia)$. By repeating the shifts in the imaginary 
direction as in (\ref{thetashift}), one then finds that
the ratio of the theta functions shows an exponential behavior 
$\varth_0(X +A)/\varth_0(X) \sim e^{\pi a t/(2KK')}$. 
Thus, along a contour  with a large radius in the world-sheet,
one of $\varphi_j,\phiinv_j$ alternatively becomes dominant.
Since this shows that the mechanism in section 4.1 works in this case,
the solution describes a surface with  four null boundaries and four cusps.
The points of the cusps in the $x_{\pm}$-plane   can be brought to
finite points as in section 4.1.

Since the Virasoro constraints are changed, the surface spanned by the solution
is not necessarily space-like anymore. This can be checked by considering the normal vector 
to the surface $N_{a} := u^{-1} \ep_{abcd} Y^{b} \del_{+} Y^{c} \del_{-} Y^{d} $, the norm
of which is
$
   N^{2} = 1 - {\kappa_{+}^{2} \kappa_{-}^{2}}/{u^{2}} \period
$
Evaluating $u = 2k^{2} \sn^{2} X + u_{0} $ in this example shows that
$N^{2} < 0$ and hence the surface is time-like.

\msection{Discussion}
We have systematically searched for the classical open string solutions 
in $AdS_{3}$ within the genus-one finite-gap solutions, 
and given a classification of the allowed solutions.
When the elliptic modulus degenerates, 
we have found a class of solutions with
six null boundaries, among which two pairs are collinear.
Adding $S^{1}$ to $AdS_{3}$,
we have also found solutions 
expressed by the elliptic  functions, which have
 four cusps and four null boundaries.

The analysis in this paper can straightforwardly be
applied to the case with the
Lorentzian world-sheet. It may also be useful for studying the classical
solutions describing the 
Wilson loops in the super Yang-Mills theory at strong coupling 
\cite{Maldacena:1998im,Rey:1998ik}. 
The classical open string solutions in $AdS_{5} \times S^{5}$ are
similarly discussed. In particular, for the strings in $AdS_{5}$, we have only to add another
pair of $\varphi_{3}, \phiinv_{3}$. In this case, these are identified 
with a complex combination of the embedding coordinates in $AdS_{5}$ as
$\varphi_{3}, \phiinv_{3} = Y_{3}\pm i Y_{4}$, and thus the solutions 
are generally (harmfully) oscillating.

In such oscillating cases, a way to remove the unwanted oscillation is
to restrict  the world-sheet, as mentioned in section 4.2. Though 
it is still non-trivial to find desired solutions with cusps and null boundaries, 
the prescription in \cite{AM2} suggests that effectively restricting the
world-sheet by conformal transformations deserves further consideration.

The essence of the solution with six null boundaries in section 4.3 is 
 the change of the sign of $\tanh B$ in front of the exponentials, which 
 ``splits'' the cusps. Similarly, more intricate behavior of the corresponding factors 
 in the higher-genus cases may produce solutions with more null boundaries.
 It is interesting to consider the relation to the mechanism provided in 
\cite{AM2}.

Most of the end points of the null segments
 in our solutions with six null boundaries are located at the infinity of the $AdS_{3}$
boundary. Since the surface is space-like, this is inevitable in the $AdS_{3}$ boundary.
However, it is desirable to bring them  to finite points in the $AdS_{5} $ boundary 
by some transformations, as discussed in \cite{AM2}.
This may be a first step 
toward  applications to the scattering amplitudes.
We would like to report progress in the analysis of the higher-genus 
finite-gap solutions, multi-cusp solutions and the applications to the scattering
amplitudes, elsewhere.

%
\setcounter{section}{0}
\appsection{Appendix}
Our  conventions of the elliptic theta functions are:
\eqb
  \theta_{ab}(w,\tau) :=
   \sum_{n=-\infty}^{\infty} \exp\Bigl[ \pi i (n+\frac{a}{2})^{2} \tau 
   + 2 \pi i (n+\frac{a}{2})(w +\frac{b}{2})\Bigr]
   \comma
\eqe
and 
\eqb
  \varth_{0}(z) := \theta_{01}(w,\tau)
  \comma \
  \varth_{1}(z) := -\theta_{11}(w,\tau)
  \comma \
  \varth_{2}(z) := \theta_{10}(w,\tau)
  \comma \
  \varth_{3}(z) := \theta_{00}(w,\tau)
  \comma
\eqe
where $w = z/(2K) $ and
$K(k)$ is the complete elliptic integral of the first kind.

In the main text, we use the formulas
\eqb
  Z(u+v)=Z(u)+Z(v)-k^2\sn\,u\,\sn\,v\,\sn(u+v) \comma\label{addZ}
\eqe
where $Z(z) := \del_{z} \ln \varth_{0}(z)$, and
\eqb
     \varth_{0}(u \pm K) &=& \varth_{3}(u) \comma \nn \\
    \varth_{0}(u \pm  iK') &=& \pm i e^{-\pi i (\pm \frac{u}{2K} + \frac{\tau}{4})}\varth_{1}(u) 
    \comma \label{thetashift}\\
    \varth_{0}(u \pm  (K+iK')) &=&  e^{-\pi i (\pm \frac{u}{2K} + \frac{\tau}{4})}\varth_{2}(u) 
    \period \nn
  \eqe

%
%
\newpage
\begin{center}
  {\bf Acknowledgments}
\end{center}

We would like to thank D. Bak, S. Hirano, N. Ishibashi, K. Ito, H. Itoyama, C. Kalousios,
T. Matsuo and K. Mohri for useful conversations. 
The work of K.S. and Y.S. is supported in part 
by Grant-in-Aid for Scientific Research from the Japan Ministry of Education, Culture, 
Sports, Science and Technology.

%
\def\thebibliography#1{\list
 {[\arabic{enumi}]}{\settowidth\labelwidth{[#1]}\leftmargin\labelwidth
  \advance\leftmargin\labelsep
  \usecounter{enumi}}
  \def\newblock{\hskip .11em plus .33em minus .07em}
  \sloppy\clubpenalty4000\widowpenalty4000
  \sfcode`\.=1000\relax}
 \let\endthebibliography=\endlist
%
%
\vspace{2ex}
\begin{center}
 {\bf References}
\end{center}
\par \vspace*{-1ex}

%

%
%
\end{document}